\documentclass[letterpaper]{article}

\def\mytitle{DIRC for a Higher Luminosity B Factory}
\usepackage{graphicx}       
\DeclareGraphicsExtensions{{.eps}}
\usepackage{xspace}

\def\mb#1{\mbox{\scriptsize #1}}
\def\G{\;\mbox{G}}
\def\GeV{\;\mbox{GeV}}
\def\MeV{\;\mbox{MeV}}

\def\ps{\;\mbox{ps}}
\def\s{\;\mbox{s}}
\def\ns{\;\mbox{ns}}
\def\nm{\;\mbox{nm}}
\def\mm{\;\mbox{mm}}
\def\m{\;\mbox{m}}
\def\cm{\;\mbox{cm}}

\def\A{\;\mbox{A}}
\def\l{\;\mbox{l}}
\def\mum{\;\mu\mbox{m}}
\def\mrad{\;\mbox{mrad}}

\def\kHz{\;\mbox{kHz}}
\def\MHz{\;\mbox{MHz}}

\def\Bbar{\kern 0.18em\overline{\kern -0.18em B}{}\xspace}
\def\Bb  {\ensuremath{\Bbar}\xspace}

\def\BaBar{{\sl B\hspace{-0.4em} {\scriptsize\sl A}\hspace{-0.4em} B\hspace{-0.4em} {\scriptsize\sl A\hspace{-0.1em}R}}}
\def\PEPII {{\sc Pep-II}}
\def\dirc{{\sc Dirc}}

\title{\mytitle}
\author{
  Thomas Hadig\\
  Stanford Linear Accelerator Center, Stanford University,
  Stanford, CA  94309\\
  for the \BaBar-\dirc\ Collaboration[1]}
\date{}

\begin{document}

\maketitle
\begin{flushright}
{\small
\vskip -7cm
hep-ex/0304028\\
SLAC--PUB--9715\\
October 2002\\}
\vskip 5.5cm
\end{flushright}

\begin{abstract}
The \dirc, a novel type of Cherenkov ring imaging device, is the primary
hadronic particle identification system for the \BaBar\ detector at the
asymmetric B-factory {\sc Pep-II} at SLAC. It is based on total internal
reflection and uses long, rectangular bars made from synthetic fused
silica as Cherenkov radiators and light guides. \BaBar\ began taking
data with colliding beams in late spring 1999. This paper describes the
challenges for the \dirc\ in a higher luminosity environment and shows
solutions to these challenges.
\end{abstract}

\section{Introduction}

The study of {\it CP}-violation using hadronic final states of the
$B\Bb$ meson system requires the ability to tag the flavor of one of
the $B$ mesons via the cascade decay $b\rightarrow c \rightarrow s,$
while fully reconstructing the final state of the other over a large
region of solid angle and momentum. The momenta of the kaons used for
flavor tagging extend up to about $2\GeV/c$, with most of them below
$1\GeV/c$. On the other hand, pions from the rare two-body decays
$B^0\rightarrow \pi^+\pi^- (K^+ \pi^-)$ must be well-separated from
kaons. They have momenta between $1.7\GeV/c$ and $4.2\GeV/c$ with a
strong momentum-polar angle correlation between the tracks (higher
momenta occur at the more forward angles because of the c.m.\ system
boost)\cite{R_PBook}. The \emph{Particle Identification} (PID) system
in \BaBar\cite{BaBarnim} is located inside the calorimeter volume.
Therefore, it should be thin and uniform in terms of radiation lengths
(to minimize degradation of the calorimeter energy resolution) and small
in the radial dimension to reduce the volume, hence, the cost of the
calorimeter.

The PID system being used in \BaBar\ is a new kind of ring-imaging
Cherenkov detector called the \dirc\cite{four} (the acronym 
\dirc\ stands for \emph{Detection of Internally Reflected Cherenkov} 
light). It is designed to be able to provide excellent $\pi/K$
separation for all tracks from $B$-meson decays from the pion Cherenkov
threshold up to $4.2\GeV/c.$ PID below $700\MeV/c$ exploits also the
$dE/dx$ measurements in the silicon vertex tracker and drift chamber.

\section{\dirc\ Concept}

\begin{figure} 
 \centerline{
  \includegraphics[width=10cm]{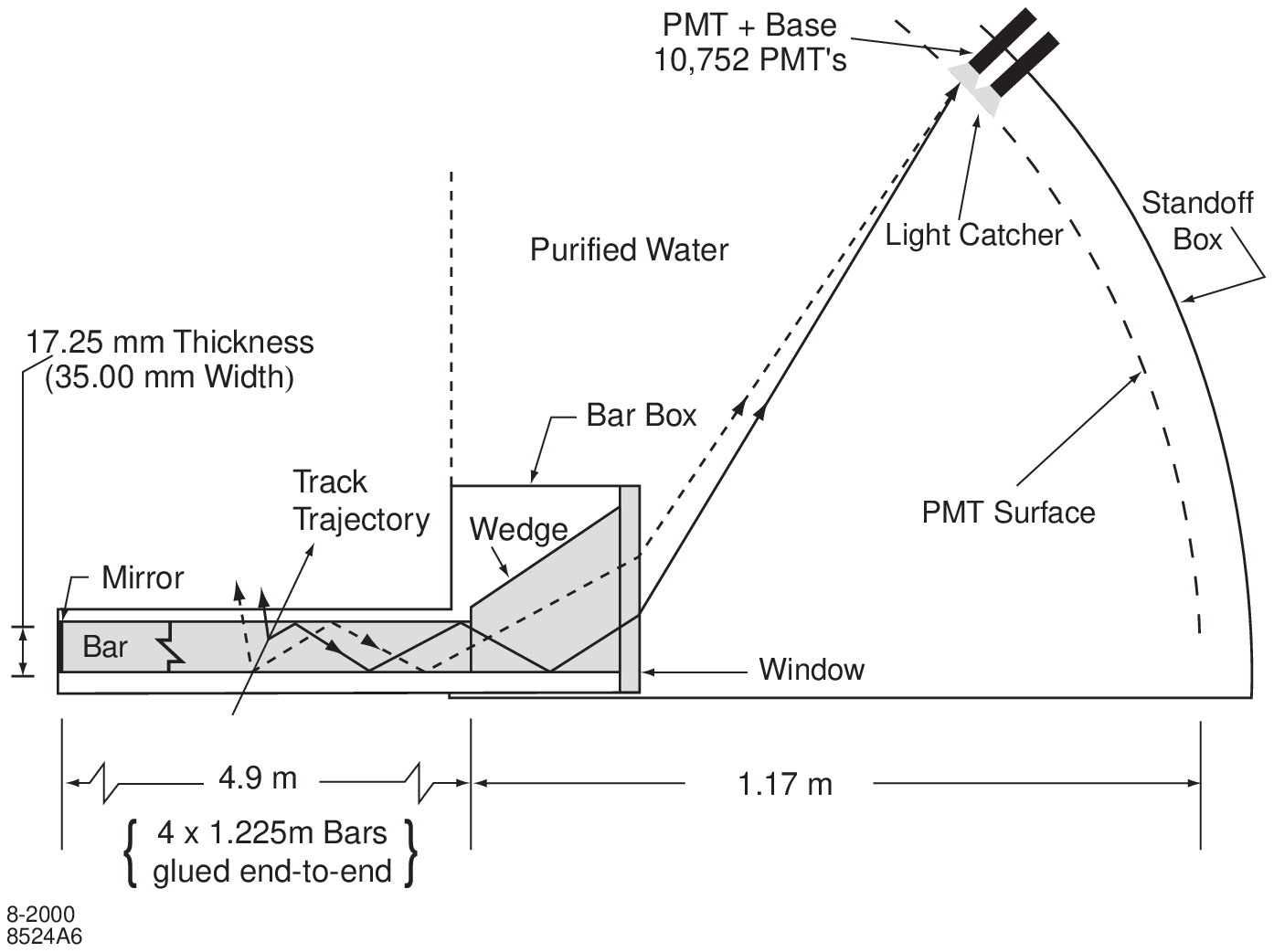}}
 \caption{\label{fig:princip}
  Schematic of the \dirc\ fused silica radiator bar and
  imaging region.
 }
\end{figure}

The \dirc\ is based on the principle that the absolute values of angles
are maintained upon reflection from a flat surface.
Figure~\ref{fig:princip} shows a schematic of the \dirc\ geometry that
illustrates the principles of light production, transport, and imaging.
The radiator material of the \dirc\ is synthetic, fused silica in the
form of long, thin bars with rectangular cross section. These bars serve
both as radiators and as light pipes for the portion of the light
trapped in the radiator by total internal reflection. Fused, synthetic
silica (Spectrosil\cite{six}) is chosen because of its resistance to
ionizing radiation, its long attenuation length, large index of
refraction, small chromatic dispersion within the wavelength acceptance
of the \dirc, and because it allows an excellent surface polish on the
bars\cite{quartzpaper}.

In the following, the variable $\theta_C$ is used to designate the
Cherenkov angle, $\phi_C$ denotes the azimuthal angle of a Cherenkov
photon around the track direction, and $n$ represents the mean index of
refraction of fused silica ($n = 1.473$) within the wavelength
acceptance of the \dirc\ ($300\nm$ to $600\nm$). The Cherenkov angle is
given by the familiar relation $\cos \theta_C = 1/(n\beta)$ with $\beta
= v/c,$ the velocity of the particle $v$ and the velocity of light $c.$

For particles with $\beta \approx 1,$ some photons will always lie
within the total internal reflection limit, and will be transported to
either one or both ends of the bar, depending on the particle incident
angle. To avoid instrumenting both ends of the bar with photon
detectors, a mirror is placed at the forward end, perpendicular to the
bar axis, to reflect incident photons to the backward, instrumented end
(see Figure~\ref{fig:princip}).

Once photons arrive at the instrumented end, most of them emerge into a
water-filled expansion region, called the \emph{standoff box}. A fused
silica \emph{wedge} at the exit of the bar reflects photons at large
angles relative to the bar axis. It thereby reduces the size of the
required detection surface and recovers those photons that would
otherwise be lost due to internal reflection at the fused silica/water
interface. The photons are detected by an array of densely packed
photomultiplier tubes (PMTs)\cite{nine,drcpmt}, each surrounded by
reflecting \emph{light catcher} cones\cite{lightcatcher} to capture
light which would otherwise miss the active area of the PMT. The PMTs
are placed at a distance of about $1.2\m$ from the end of the bars. The
expected Cherenkov light pattern at this surface is essentially a conic
section, where the cone opening-angle is the Cherenkov production angle
modified by refraction at the exit from the fused silica window.

\begin{figure} 
 \centerline{
  \includegraphics[width=7.5cm]{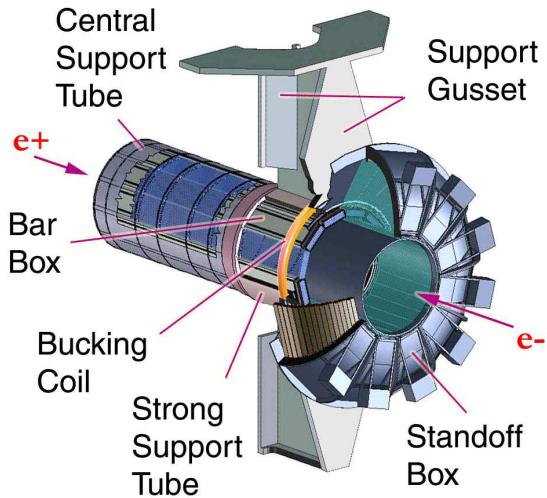}}
 \caption{\label{fig:mechelmsa}
  Exploded view of the \dirc\ mechanical support structure. The iron
  magnetic shield is not shown.
 }
\end{figure}

The \dirc\ bars are arranged in a 12-sided polygonal barrel (see
Figure~\ref{fig:mechelmsa}). Because of the beam energy asymmetry (at
the $\Upsilon (4S)$, {\sc Pep-II} collides $9\GeV$ electrons on
$3.1\GeV$ positrons), particles are produced preferentially forward in
the laboratory. To minimize interference with other detector systems in
the forward region, the \dirc\ photon detector is placed at the backward
end. The bars are placed into 12 hermetically sealed containers, called
\emph{bar boxes}, made of aluminum-hexcel panels. Dry nitrogen gas flows 
through each box, and is monitored for humidity to ensure that the bar
box to water interface remains sealed. Each bar box contains 12 bars,
for a total of 144 bars. Within a bar box the 12 bars are optically
isolated by a $\sim 150\mum$ air gap between neighboring bars, enforced
by custom shims made from aluminum foil.

The bars are $17\mm$ thick, $35\mm$ wide, and $4.9\m$ long. Each bar is
assembled from four $1.225\m$ pieces that are glued end-to-end; this
length is the longest high-quality bar currently
obtainable\cite{quartzpaper,eight}. The bars are supported at $600\mm$
intervals by small nylon buttons for optical isolation from the bar box.
Each bar has a fused silica wedge glued to it at the readout end. The
wedge, made of the same material as the bar, is $91\mm$ long with very
nearly the same width ($33\mm$) as the bars and a trapezoidal profile
($27\mm$ high at bar end, and $79\mm$ high at the light exit end). The
bottom of the wedge (see Figure~\ref{fig:princip}) has a slight ($\sim
6\mrad$) upward slope to minimize the displacement of the downward
reflected image due to the finite bar thickness. The 12 wedges in a bar
box are glued to a common $10\mm$ thick fused silica window which
provides the interface and seal to the purified water in the standoff
box.

The standoff box (see Figure~\ref{fig:mechelmsa}) is made of stainless
steel, consisting of a cone, cylinder, and 12 sectors of PMTs. It
contains about $6000\l$ of purified water. Water is used to fill this
region because it is inexpensive and has an average index of refraction
($n \approx 1.346$) reasonably close to that of fused silica, thus
minimizing the total internal reflection at silica-water interface.
Furthermore, its chromaticity index is a close match to that of fused
silica, effectively eliminating dispersion at the silica-water
interface. The iron gusset supports the standoff box. An iron shield,
supplemented by a \emph{bucking coil}, surrounds the standoff box to
reduce the field in the PMT region to below $1\G$\cite{bfield}.

The PMTs at the rear of the standoff box lie on an approximately
toroidal surface. The distance from the end of the bar to the PMTs is
$\sim 1.17\m.$ Each of the 12 PMT sectors contains 896 PMTs (ETL model
9125\cite{nine,drcpmt}) with $29\mm$ diameter, in a closely packed array
inside the water volume. A double o-ring water seal is made between the
PMTs and the vessel wall. The PMTs are installed from the inside of the
standoff box and connected via a feed-through to a base mounted outside.
The hexagonal light catcher cone is mounted in front of the photocathode
of each PMT which results in an effective active surface area light
collection fraction of about 90\%.

The \dirc\ occupies $80\mm$ of radial space in the central detector
volume including supports and construction tolerances with a total of
about 19\%\ radiation length thickness at normal incidence. The radiator
bars cover a solid angle corresponding to about 94\%\ of the azimuth
and 83\%\ of the c.m.\ polar angle cosine.


\section{\dirc\ Performance}

The performance of the \dirc\ is influenced by two factors. First,
information from the PMTs has to be separated into background and
signal. Second, the $\theta_C$ resolution of the signal has to allow for
the separation of different particle species.

\begin{figure*} 
 \centerline{
  \includegraphics[width=7.5 cm]{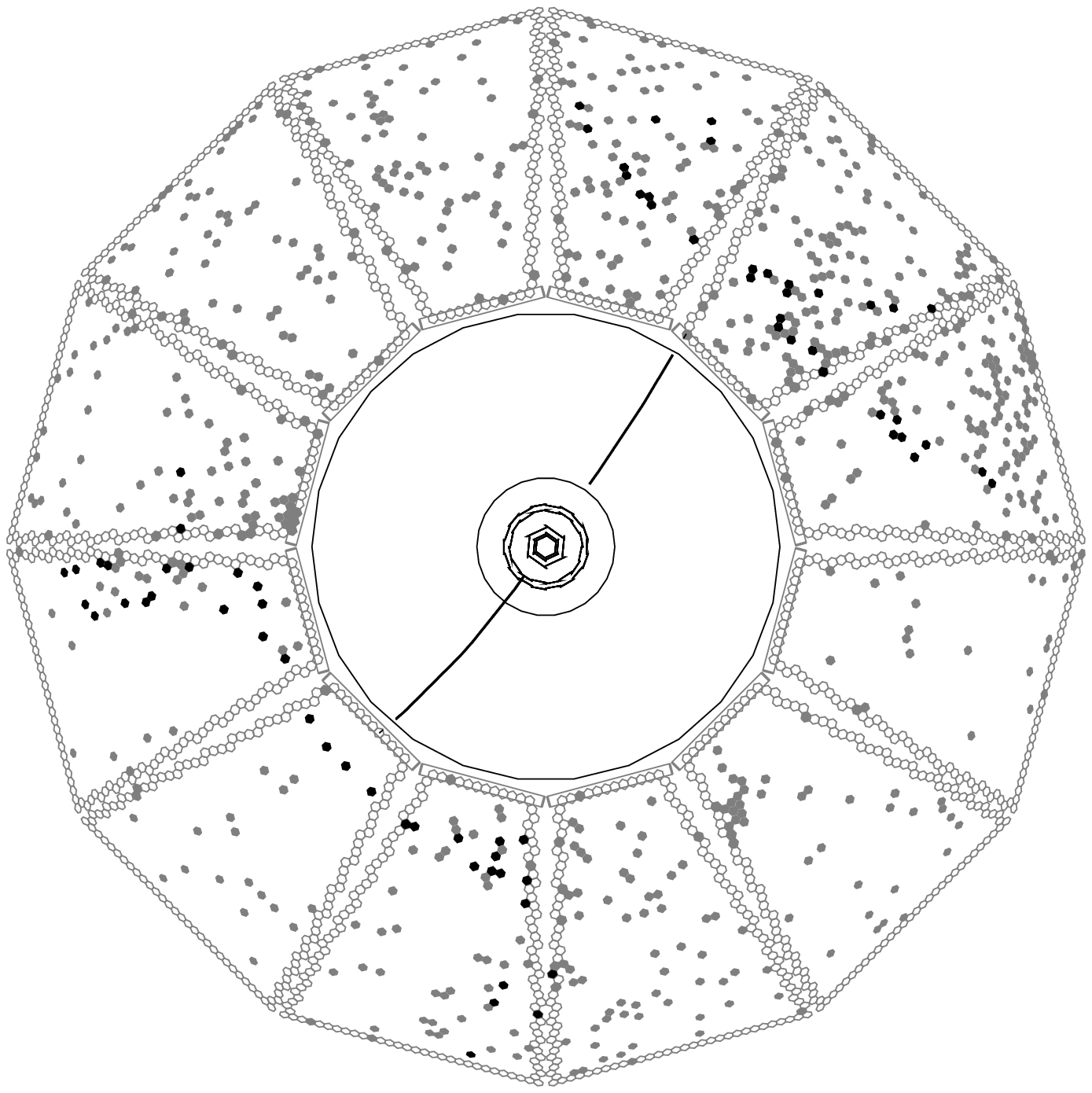}
  \includegraphics[width=7.5 cm]{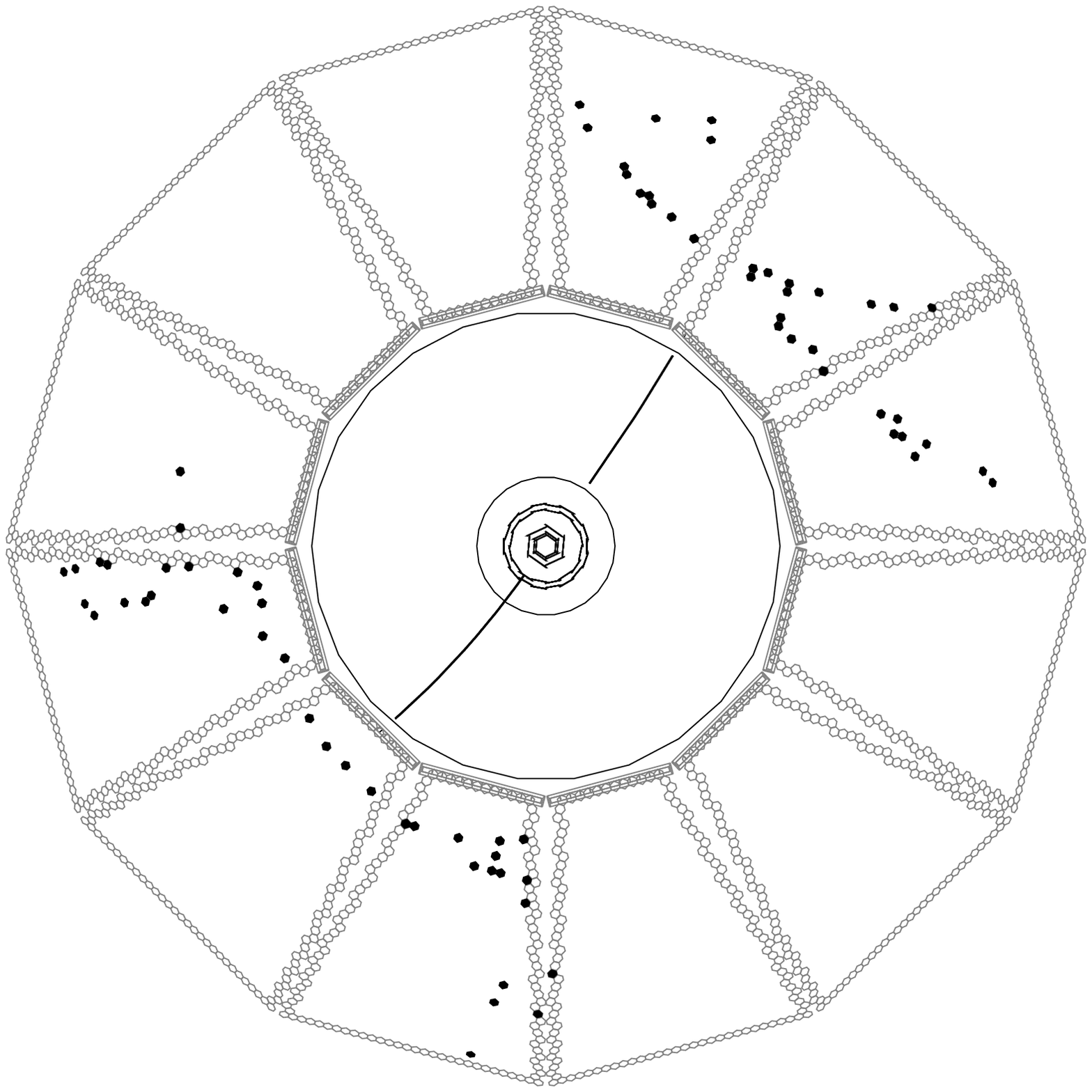}}
 \caption{\label{fig:event}
  Display of an $e^+e^-\rightarrow\mu^+\mu^-$ event
  reconstructed in \BaBar\ with two different time cuts. On the left,
  all \dirc\ PMTs with signals within the $\pm 300\ns$ trigger window
  are shown. On the right, only those PMTs with signals within $8\ns$ of
  the expected Cherenkov photon arrival time are displayed.}
\end{figure*}

Figure~\ref{fig:event} shows a typical di-muon event ($e^{+}e^{-}
\rightarrow \mu^{+}\mu^{-}$). In addition to the signals caused by 
Cherenkov light from the muon tracks, about 500 background signals can
be seen in the $600\ns$ readout window centered around the trigger. This
background is dominated by low energy photons from the
\PEPII\ machine hitting the standoff box.

\begin{figure}
 \centerline{
  \includegraphics[width=7.5cm]{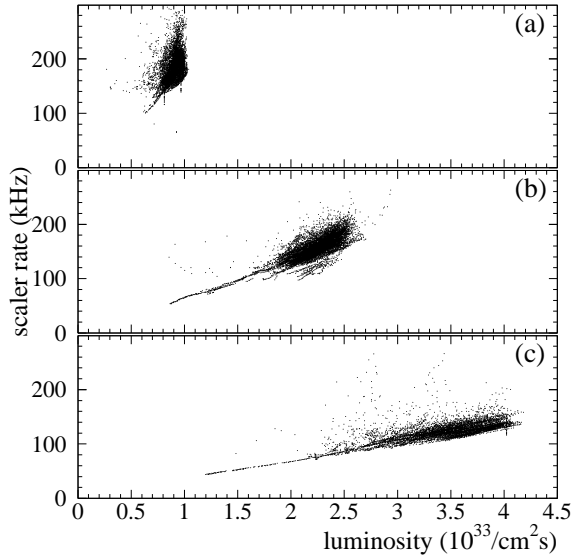}}
 \caption{\label{fig:scaler}
  Maximum scaler rates measured for three different accelerator and 
  shielding configurations in March 2000 (a), October 2000 (b) and
  August 2001 (c). 
 }
\end{figure}

The time-to-digital converter (TDC) chip\cite{tdc} used in the 
\dirc\ data readout is designed such that a dead time of about 5\% occurs 
at an input rate of $250\kHz.$ Some care in machine tuning is required
to stay under a limit of $250\kHz$ per tube. To monitor this rate, one
PMT in each sector is read out via a scaler. Figure~\ref{fig:scaler}
shows the maximum scaler rate as a function of the \PEPII\ luminosity
during data taking in 2000 and 2001. In March 2000 the accelerator
operated at a peak luminosity of $10^{33}\cm^{-2}\s^{-1}$.
Figure~\ref{fig:scaler}(a) shows, at a value corresponding to only one
third of the design luminosity, the PMT rates reached a level that
caused noticeable dead times. Due to those findings, lead shielding was
installed in the summer of 2000 around the beam line components near the
backward end-cap. Initially, localized shielding was added in the form
of lead bricks which were stacked around the beam pipe and in front of a
large quadrupole. This shielding significantly improved the background
situation so that noticeable TDC dead times were reached only at $2.5
\times 10^{33}\cm^{-2}\s^{-1}$ (Figure~\ref{fig:scaler}(b)). During the 
shutdown in January 2001, the localized lead brick shielding was
replaced by an engineered, homogeneous lead shielding of $5\cm$ to
$8\cm$ thickness covering the inside radius of the standoff box and is
easily removable to facilitate access to the central detector and beam
line components. As is shown in figure~\ref{fig:scaler}(c), the maximum
scaler rates at luminosities of $4.2 \times 10^{33}/\cm^{-2}\s^{-1}$ are
well below the level that would cause TDC inefficiencies.

During the shutdown following the 2001-2002 run, we plan to replace the
TDCs with a faster version with deeper buffering that is designed to
have a deadtime of less than 5\% at $2.5\MHz$ input rate, suitable for
luminosities of at least $10^{34}\cm^{-2}\s^{-1}.$

The \dirc\ is intrinsically a three-dimensional imaging device. Photons
are focused onto the phototube detection surface via a ``pinhole''
defined by the exit aperture of the bar, so that the photon propagation
angles $\alpha_x$ and $\alpha_y$ can be measured in two-dimensional
space, where $x$ (bar width) and $y$ (bar thickness) are the directions
transverse to the bar axis. The travel time of the photon down the bar
is also related to the photon propagation angle ($\alpha_z$) with
respect to the bar axis. Imaging in the \dirc\ occurs in all three of
these dimensions, by recording the time of the PMT signal. As the track
position and angles are known from the tracking system, these three
$\alpha$ angles can be used to determine the two angles ($\theta_C,
\phi_C$). This over-constraint on the angles is particularly
useful in dealing with ambiguities and high background rates.

The single photon resolution can be calculated from the geometrical, the
chromatic and the transport term. The geometrical uncertainty is, for
the pinhole optics of the \dirc, given by the standoff box size, the
bar and the PMT size. With $7.2\mrad,$ it is the single biggest
contribution to the single photon resolution. The chromatic uncertainty
originates from the fact that the photons are produced with different
wavelengths and, therefore, have different cone opening angles
$\theta_C.$ Using the range in photon wavelength accepted by the PMTs,
the chromatic effect contributes to the resolution with $5.4\mrad.$ The
final uncertainty comes from imperfections of the bars and the glue
joints. This transport term is approximately $2\mrad$ to $3\mrad.$

\begin{figure}
 \centerline{
  \includegraphics[width=7.5cm]{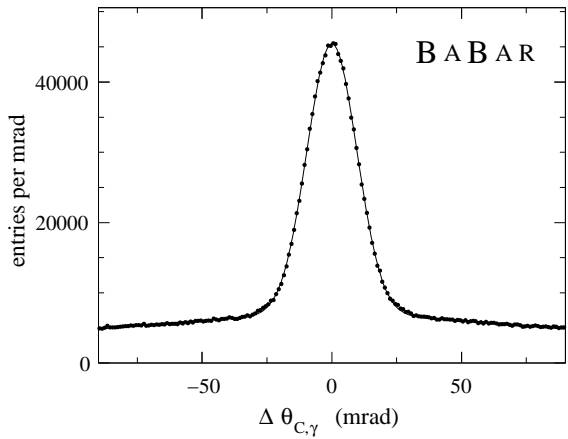}
 }
 \caption{\label{fig:singres}
  The difference between the measured and
  expected Cherenkov angle for single photons, $\Delta \theta_{C,\gamma}$
  for single muons in $\mu^+\mu^-$ events.
  The curve shows the result of a fit of two Gaussians to the data.
  The width of the narrow Gaussian is $9.6\mrad.$}
\end{figure}

Combining the three contributions, a single photon resolution of
$\sigma_{\theta_{C},\gamma} = 9.5\mrad$ is expected for the \dirc.
Figure~\ref{fig:singres} shows the single photon angular resolution
obtained from di-muon events. There is a broad background of less than
10\%\ relative height under the peak that originates mostly from
track-associated sources, such as $\delta$ rays, reflections off the
glue-fused silica boundaries, and combinatorial
background\cite{jerrykevin}. The width of the peak translates to a
resolution of about $9.6\mrad,$ in good agreement with the expected
value.

The resolution on the track Cherenkov angle is defined by a correlated
term and the single photon resolution which scales with the number of
photoelectrons:
\begin{equation}
	\sigma_{\theta_{C}} = \sigma_{\theta_{C},\mb{corr}} \oplus 
           {\sigma_{\theta_{C},\gamma}\over \sqrt{N_{\mb{pe}}}}
\end{equation}
where $N_{pe}$ is the number of photons detected per track.

\begin{figure} 
 \centerline{
  \includegraphics[width=7.5cm]{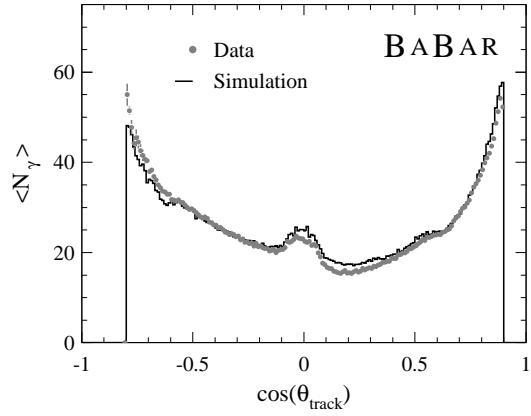}
 }
 \caption{\label{fig:nphot}
  Number of detected photons versus track polar angle
  for reconstructed tracks in di-muon events compared to Monte Carlo
  simulation.
  The mean number of photons in the simulation has been tuned to match
  the data.}
\end{figure}

The number of photoelectrons varies between 16 for small values of
$\cos\theta_{\mb{track}}$ at the center of the barrel and 60 at large
values of $\cos\theta_{\mb{track}}$ as is shown in
Figure~\ref{fig:nphot}. This variation is well reproduced by Monte Carlo
simulation and can be understood from the geometry of the \dirc. The
number of Cherenkov photons varies with the path length of the track in
the radiator, it is smallest at perpendicular incidence at the center
and increases towards the ends of the bars. In addition, the fraction of
photons trapped by total internal reflection rises with larger values of
$\cos\theta_{\mb{track}}.$ This increase in the number of photons for
forward going tracks is a good match to the increase in performance
required at larger momentum.

\begin{figure}
 \centerline{
  \includegraphics[width=7.5cm]{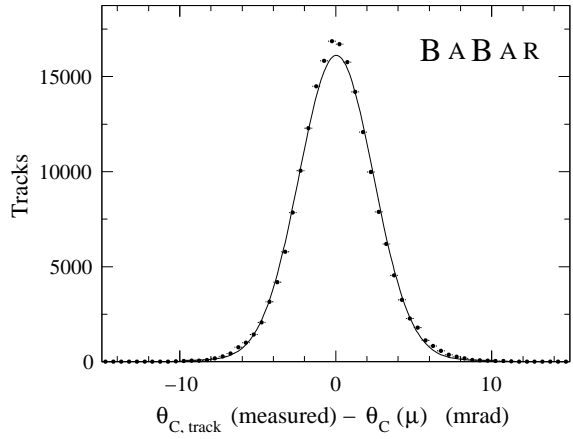}}
 \caption{\label{fig:trackres}
  Resolution of the reconstructed Cherenkov polar angle per track for
  di-muon events.
  The curve shows the result of a Gaussian fit with a resolution of $2.4\mrad.$}
\end{figure}

With the present alignment, the track Cherenkov angle resolution for
di-muon events is shown in Figure~\ref{fig:trackres}. 
The width of the fitted Gaussian distribution is $2.4\mrad$ compared to
the design goal of $2.2\mrad.$

The measured time resolution is $1.7\ns,$ close to the intrinsic $1.5\ns$
transit time spread of the PMTs. This resolution is used to efficiently 
distinguish background from signal photons but is not sufficient to
improve the Cherenkov angle resolution. 

\begin{figure} 
 \centerline{
  \includegraphics[width=7.5cm]{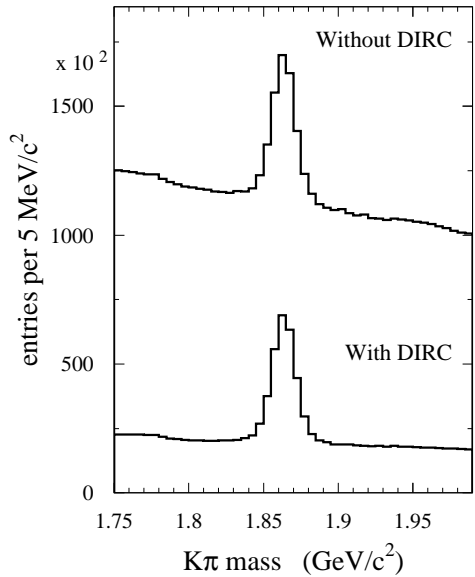}}
 \caption{\label{fig:kpi}
  Invariant K$\pi$ inclusive mass spectrum  with and without the
  use of the \dirc\ for kaon identification. The mass peak corresponds
  to the decay of the $D^0$ particle.}
\end{figure}

Figure~\ref{fig:kpi} shows an example of the use of the \dirc\
for particle identification.
The $K\pi$ invariant mass spectra are shown with and
without the use of the \dirc\ for kaon identification.
The peak corresponds to the decay of the $D^0$ particle.

The PID performance of the \dirc\ has been studied with a sample of 
pions and kaons, selected kinematically using 
$D^0 \rightarrow K^-\pi^+$ decays from inclusive $D^{*}$ production.

\begin{figure} 
 \centerline{
  \includegraphics[width=7.5cm]{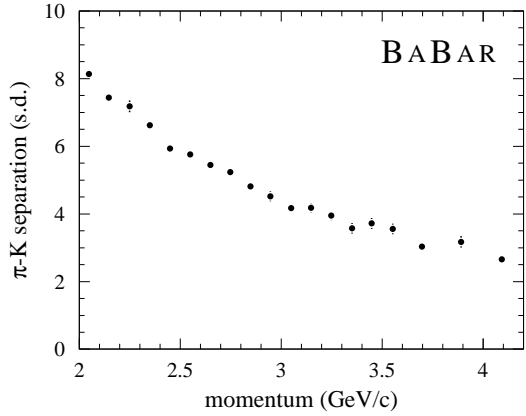}}
 \caption{\label{fig:sep}
  \dirc\ $\pi/K$ separation versus track momentum measured in 
  $D^0 \rightarrow K^-\pi^+$ decays selected kinematically from 
  inclusive $D^{*}$ production.}
\end{figure}

The $\pi/K$ separation power of the \dirc\ was defined as the
difference of the expected Cherenkov angles for pions and kaons,
divided by the measured track Cherenkov angle resolution.
As shown in Figure~\ref{fig:sep}, the separation between kaons and
pions at $3\GeV/c$ is about $4.4\sigma$, within 10\% of the design goal.

\begin{figure} 
\centerline{\includegraphics[width=7.5cm]{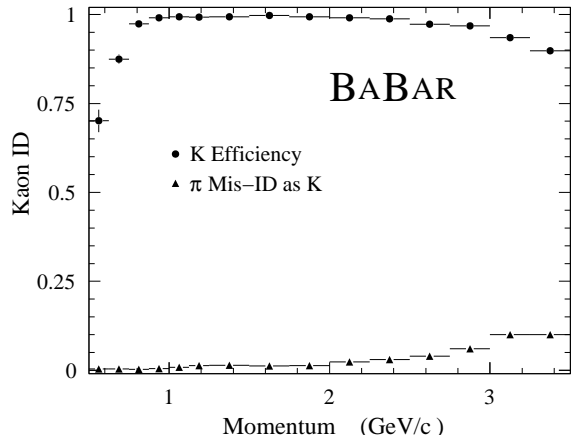}}
\caption{Efficiency and misidentification probability for the
selection of charged kaons as a function of track momentum, 
for a particular choice of particle selection criteria.
The data use $D^0 \rightarrow K^-\pi^+$ decays selected kinematically from
inclusive $D^{*}$ production.}
\label{fig:kpi_eff}
\end{figure}

The efficiency for correctly identifying a charged kaon
that traverses a radiator bar and the probability to wrongly identify
a pion as a kaon are also determined from the inclusive $D^{*}$ sample
 and are shown as a function of the track
momentum in Figure~\ref{fig:kpi_eff} for a particular choice of
particle selection criteria.
The kaon selection efficiency and pion misidentification, integrated
over the 
$K$ and $\pi$ momentum spectra of the $D^{*}$ control sample, 
are $97.97 \pm 0.07$\% (stat. only) and $1.83 \pm 0.06$\% (stat. only),
respectively.

A report on the operational experience with the \dirc\ detector can
be found in \cite{ieee2001}.


\section{\dirc\ in Higher Luminosity Environment}

In a higher luminosity environment, the accelerator-induced background
will cause the main challenge. From a set of special runs taken in
February 2002, the background rate as a function of the beam currents
$I_{\mb{HER}}$ and $I_{\mb{LER}}$ in the high and low energy ring,
respectively, and the luminosity ${\cal L}$ can be parameterized as
\begin{equation}
  R = 13 {\kHz\over\A} I_{\mb{HER}} + 18 {\kHz\over\A} I_{\mb{LER}} +
  10 {\kHz\over 10^{33}\cm^{-2}\s^{-1}} {\cal L}.
\end{equation}
For the conditions at the beginning of this year, the current terms
dominate but for higher luminosities, the corresponding term will
contribute significantly. The anticipated limit of the luminosity
\PEPII\ will be able to provide without major upgrade is
$4 \times 10^{34}\cm^{-2}\s^{-1}.$ With some upgrades to the
shielding and the TDC upgrade, the \dirc\ will be able to reliably work
in this environment. At higher luminosities, such as expected after
major \PEPII\ updates or for Super\BaBar, more changes will be
necessary.

A solution to high background rates is to reduce the size of the
standoff box. However, without modifications to the design, this
would lead to an increase in the geometric resolution term and would
cause a considerable decrease in the \dirc\ performance.

\begin{figure}
 \centerline{
  \includegraphics[width=.6\hsize]{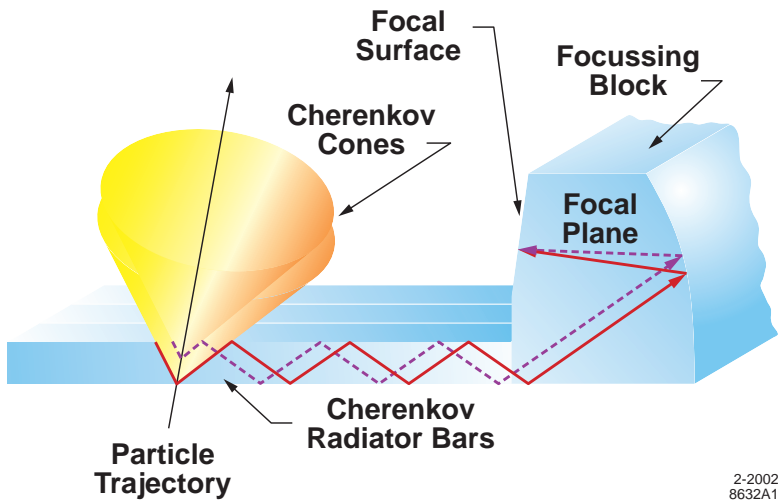}}
 \caption{\label{fig:newdesign}
  New design of \dirc\ standoff region using focusing optics.
 }
\end{figure}

A modified design, as shown in figure~\ref{fig:newdesign}, employs
smaller sized photodetectors as well as focusing optics effectively
removing the bar size uncertainty in the focusing plane. These two
enhancements can be balanced so that the size of the standoff region
can be reduced while keeping a similar or improved geometrical resolution.

\begin{figure}
 \centerline{
  \includegraphics[width=.6\hsize]{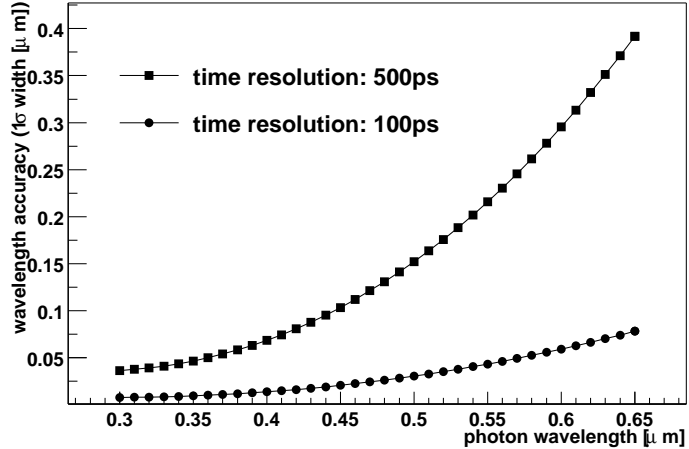}}
 \caption{\label{fig:wavelengthtime}
  Achievable wavelength resolution as a function of photon wavelength
  assuming $100\ps$ and $500\ps$ timing resolution.
 }
\end{figure}

In addition, smaller photodetectors with a better transit time spread
will improve the timing resolution of the detector. This can be used to
further reduce the background by using tighter time cuts and to reduce
the chromatic uncertainty. The travel time of a photon is the difference
between the time the particle hits the bar and the arrival time of the
photon. It depends on the path length and the group velocity in the
transit medium. The group velocity itself depends on the refractive
index. Thus, the travel time and path length yield a measure of the
wavelength of the photon. Figure~\ref{fig:wavelengthtime} shows a
calculation demonstrating that a timing resolution in the order of
$100\ps$ is needed for a significant improvement in the chromatic
uncertainty.

Reducing the range of wavelength accepted in the detection system will
reduce the range of Cherenkov cone opening angles contributing and,
thus, reduce the chromatic uncertainty. The drawback of this solution is
the loss in the number of photons detected per track causing a reduced
suppression of the single photon resolution term. These two effects can
be balanced in order to reach the optimal total $\theta_C$ resolution.

With a timing resolution in the order of $100\ps,$ the ratio of the
track to detector distance over the path length will provide an
additional measure of the photon angle. In some regions of phase space,
especially at small dip angles, its resolution will be competitive to
the angle resolution from the position of the PMT hit.

A more detailed discussion can be found in \cite{Ratcliff:2001ss}.

One of the photon detector candidates is the flat panel PMT H-8500 by
Hamamatsu\cite{Hamamatsu}. It has $8 \times 8$ square pads on a size of
$50\mm\times 50\mm.$ This leads to a high area efficiency and packaging
density. The timing resolution is expected to be in the range of
$100\ps$ to $200\ps.$ The disadvantage of the currently available model
is its low gain in the order of $1.6 \times 10^6.$

\begin{figure}
 \centerline{
  \includegraphics[width=.6\hsize]{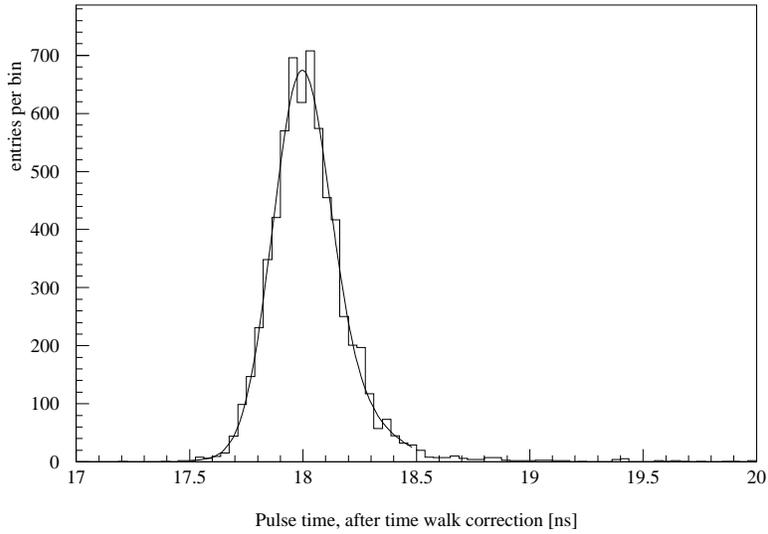}}
 \caption{\label{fig:pmttiming}
  Time resolution of the Hamamatsu H-8500 PMT. The data was recorded
  using two different thresholds to allow for time walk correction. No
  other correction has been applied.
 }
\end{figure}

Studies of the timing resolution on an early pre-production version were
encouraging. The time was measured at two thresholds for each signal in
order to perform a time walk correction. The corrected time
distribution\cite{jerryrich} is shown in figure~\ref{fig:pmttiming}. A
fit to a double Gaussian plus a second order polynomial describes the
data well. The narrow Gaussian has a width of $125\ps.$ This value is
not yet corrected for the light pulse width and instrumentational
effects. In addition, the double threshold method limits the data to PMT
signals with a pulse height above the higher threshold. This removes
some fraction of low signal events.

Currently, studies of the efficiency uniformity of the PMT area are
in process and results will be available in near future.


\section{Improvements for Future \dirc\ Detectors}

The design described in the previous section will be sufficient to
provide the required resolution of the \dirc\ for possible upgrade
scenarios. However, for the Super\BaBar\ environment with a design
luminosity of $10^{36}\cm^{-2}\s^{-1},$ the projected occupancy in the
tracking system makes building a completely new detector necessary.

This allows to reconsider the radiator and light guide design.
Synthetic fused silica is still the material of choice as it fulfills
all requirements in terms of transparency, radiation hardness, 
achievable uniformity in material and optical finish. In addition,
it has a small radiation length which is benifical for a calorimeter
outside of the \dirc\ system.

Increasing the thickness of the radiators will lead to an increased
number of photons per track but it will also increase the amount of
material in front of the calorimeter, so the thickness of the radiator
can not be changed significantly.

\begin{figure}
 \centerline{
  \includegraphics[width=.6\hsize]{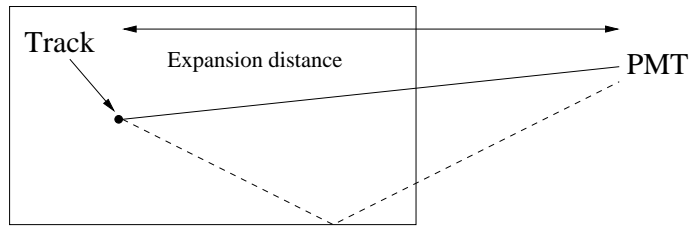}}
 \caption{\label{fig:plategeo}
  Plate geometry.
 }
\end{figure}

Increasing the width of the radiators, such as using one plate instead
of several bars, has advantages in the geometrical term of the single
photon resolution. In a plate geometry, there are less side bounces and
the path length for a photon measured in a given PMT with a known track
position changes drastically with each additional side bounce.
Therefore, the hit time can be used to determine the number of side
bounces and because of this the exit position of the photon at the
instrumented bar end can be inferred. This is shown in
figure~\ref{fig:plategeo}, where the full line and the dashed line have
significantly different path lengths.

There are a few examples in the following on the expected resolution for
the different setups. A detailed discussion of the geometries can be
found in \cite{Ratcliff:2001ss}. Please note that the resolutions are
given in terms of the photon exit angles $\alpha$ at the end of the bar
or plate. These are different to the Cherenkov angle resolution
$\sigma_{\theta_C}.$ However, there is a strong correlation between
$\sigma_{\alpha_y}$ and $\sigma_{\theta_C}$ while the correlation between
$\sigma_{\alpha_x}$ and $\sigma_{\theta_C}$ is weaker.

\bigskip
\noindent {\bf Current \BaBar\ \dirc}

The current \dirc\ setup has round PMTs with $29\mm$ diameter and rectangular
bars of size $35\mm \times 17.5\mm.$ The resolution on the photon exit angle
at the edge of the bar is
\begin{equation}
  \sigma_{\alpha_x} 
  \approx {1\over L_{\mb{SOB}} } 
          \sqrt{\sigma_x^2(\mbox{bar}) + \sigma_x^2(\mbox{pixel})}
  \approx {1\over 1170\mm } 
          \sqrt{ {1\over 12}(35\mm)^2  + {1\over 16}(29\mm)^2}
  = 10.6\mrad
\end{equation}
and
\begin{equation}
  \sigma_{\alpha_y} \approx 7.6\mrad
\end{equation}

The interface fused silica --- water magnifies the angles thus improving
the resolution by approximately 10\% leading to $\sigma_{\alpha_{x/y}} =
9.5\mrad/6.9\mrad.$

\bigskip
\noindent {\bf Focusing optics for \BaBar\ like \dirc}

A focusing optic in $y$ with a pixel size of $6\mm\times 6\mm$ and a 
corresponding standoff distance of $L=250\mm$ leads to
\begin{equation}
  \sigma_{\alpha_y} 
  \approx {1\over L} 
          \sqrt{\sigma_x^2(\mbox{pixel})}
  \approx {1\over 250\mm } 
          \sqrt{ {1\over 12}(6\mm)^2}
  = 6.9\mrad
\end{equation}
similar to the current $\alpha_y$ resolution. The resolution in $x$ will
get worse as focusing in both dimensions would complicate ambiguity
resolution. This can be compensated by the reduced chromatic
uncertainty.

\bigskip
\noindent {\bf Focusing optics with plates for Super\BaBar}

In order to improve the resolution, a smaller pad size of $2\mm$ in $y$
direction and $6\mm$ in $x$ direction can be used. The resolution can
be described by
\begin{equation}
  \sigma_{\alpha_y} 
  \approx {1\over L} 
          \sqrt{\sigma_x^2(\mbox{pixel})}
  \approx {1\over 250\mm } 
          \sqrt{ {1\over 12}(2\mm)^2}
  = 2.3\mrad
\end{equation}
in $y$ direction and in $x$
\begin{equation}
  \sigma_{\alpha_x} 
  \approx {1\over L} 
          \sqrt{\sigma_x^2(\mbox{track}) + \sigma_x^2(\mbox{pixel})}
  \approx {1\over 1000\mm } 
          \sqrt{ (4\mm)^2 + {1\over 12}(6\mm)^2}
  = 4.3\mrad
\end{equation}
In the last equation $L$ is the length of the region where the photon
can expand without side bounces. This length varies depending on the
number and positions of the side bounces.

Together with the improved chromatic resolution, this leads to a 
significant improvement in the single photon resolution compared to
the current \BaBar\ \dirc.

The major uncertainty not directly related to the \dirc\ design is the
correlated uncertainty on the position of the track at the entry into
the radiator. In order to improve the particle separation, it is not
sufficient to improve the intrinsic \dirc\ uncertainty only. A possible
solution is to add a tracking device outside the \dirc\ volume. This
would also allow for finding particle conversions and decays inside the
radiator material.

A further improvement is to enlarge the angular coverage of the particle
identification system by using an end-cap device. The \dirc\ covers the
barrel region only. Additional challenges for an end-cap device are the
limited amount of space in the end-cap, the higher radiation levels
close to the beam-pipe, and that the readout has to be done inside the
magnetic field which makes standard PMTs unusable. Interesting
alternative photodetectors have been suggested but they still need to be
studied in detail before an evaluation on the feasibility of an end-cap
\dirc\ can be performed.


\section{Conclusions}

The \dirc\ is a novel particle identification system used for the first
time in the \BaBar\ environment. It is performing close to the design
and has a significant influence on most physics analysis.

Increasing the luminosity beyond the \BaBar\ design values will result
in challenges for the \dirc\ system mainly due to increased background
levels of low energy photons from \PEPII. Improvements in shielding and
in the maximal data taking rate will allow for running the current
detector up to the anticipated \PEPII\ limits. Beyond that, reducing the
size of the standoff box will be necessary. In order to keep the
Cherenkov angle resolution on the design level, the standoff box needs
to be replaced by focusing optics with smaller sized photodetectors.
Timing resolution in the order of $100\ps$ is vital to reduce the
chromatic uncertainty and can effectively separate signal from
background photons. Research on candidate devices has started and first
tests have shown promising results.

At luminosities in the order of $10^{36} \cm^{-2}\s^{-1}$ as envisioned
for Super\BaBar, a completely new \dirc\ can be built. Fused silica
plates instead of bars and even smaller PMT pad sizes along with an
improved tracking will allow for significant improvements in the
Cherenkov angle resolution.

Enlarging the angular coverage of a \dirc\ particle identification
system will require an end-cap device. The challenges for this device
include photodetection inside of the magnetic field and the limited
amount of space prohibiting large expansion volumes.




\end{document}